\documentclass[11pt,twoside]{article}

\usepackage{asp2006}
\usepackage{epsf}
\usepackage{psfig}
\usepackage{lscape}

\begin{document}
\title{The Long Faint Tail of the High-Redshift Galaxy Population}   
\author{
M.\ Sawicki\altaffilmark{1,2},
I.\ Iwata\altaffilmark{3}, 
K.\ Ohta\altaffilmark{2},  
D.\ Thompson\altaffilmark{4},
N.\ Tamura\altaffilmark{5},
\\
M.\ Akiyama\altaffilmark{5},
K.\ Aoki\altaffilmark{5}, 
M.\ Ando\altaffilmark{2},
and
G.\ Kiuchi\altaffilmark{2}
}

\altaffiltext{1}{Physics Department, University of California, Santa Barbara, CA 93106, USA}

\altaffiltext{2}{Department of Astronomy, Kyoto University, Sakyo-ku, Kyoto, 606-8502, Japan}

\altaffiltext{3}{Okayama Astrophysical Observatory, NAOJ, Okayama, 719-0232, Japan}

\altaffiltext{4}{LBT Observatory, University of Arizona, 933 N.\ Cherry Avenue, Tucson, AZ 85721-0065, USA}

\altaffiltext{5}{Subaru Telescope, NAOJ, 650 North A`ohoku Place, Hilo, HI 96720, USA}

\begin{abstract} 
We study the properties of very faint, sub-$L^*$ Lyman break galaxies
at $z$$\sim$2--5 --- thus far a largely neglected but numerically and
energetically very important population.  We find that the LBG
luminosity function undergoes luminosity-dependent evolution: the
number of luminous galaxies remains constant while the number of faint
ones grows with time.  The total UV luminosity density increases with
cosmic time from at least $z$$\sim$5 until reaching a peak or a
plateau around $z$$\sim$2 --- behaviour that is governed by the
sub-$L^*$ galaxies in the LF's ``faint tail''.  Using broadband SED
fitting we find a nearly-linear relationship between SFR and galaxy
stellar mass at $z$$\sim$2. A typical $L^*$ LBG at $z$$\sim$2 shows a
stellar mass of $\sim$10$^{10}$$M_\odot$, remarkably similar to the
bimodality mass at low redshift. This similarity suggests that the
mechanisms responsible for the galaxy bimodality at low-$z$ may have
also been at play at $z$$\sim$2.
\end{abstract}



\section{Introduction}

Until very recently, studies of $z$$>$1 galaxies have focused
primarily on luminous, vigorously star-forming objects such as
submillimetre sources or $L$$\sim$$L^*$ LBGs that are forming stars at
rates of 10s or 100s $M_\odot$/yr.  Such studies have largely
neglected the less glamorous but far more numerous faint, sub-$L^*$
galaxies that are forming stars at much lower rates. Although
individually faint, these sub-$L^*$ objects are very numerous and so
collectively they generate more than half the total UV luminosity
density of the universe at high redshift. They are, thus, extremely
important contributors to the story of star formation and metal
enrichment in the early Universe.  Moreover, the presence of a break
at $L^*$ in the galaxy LF suggests that galaxies below $L^*$ differ
substantially from those above it. Because of this, our understanding
of the mechanisms that drive galaxy evolution at high redshift may
profit from comparing the properties of the better-studied luminous
($L$~$^>_\sim$~$L^*$) objects with those of the neglected sub-$L^*$
galaxies.  This paper summarizes the results of our recent work aimed
at understanding these hitherto neglected, individually modest but
collectively very important objects.  They are the long faint tail of
the high-$z$ galaxy population.

\section{Luminosity function and luminosity density}

\subsection{Data}

The shape of the galaxy LF bears the imprint of galaxy formation and
evolution processes.  We study the global statistics of
sub-$L^*_{z=3}$ galaxies at $z$$\sim$2, 3, 4, and 5 using two deep,
large-area, multi-field imaging surveys.

At $z$$\sim$2, 3, and 4 we use the Keck Deep Fields (KDF) of
Sawicki \& Thompson (2005, 2006a, b).  The KDF use the {\it very same}
$U_n G {\cal R} I$ filter set and color-color selection criteria as
are used in the well-known work of
\citet{sawicki:steidel1999, sawicki:steidel2003, sawicki:steidel2004}.
However, in contrast to the Steidel et al.\ work, the KDF reach ${\cal
R}_{lim}(AB)$=27, which is 1.5 mag deeper than the Steidel et al.\
surveys and significantly below $L^*_{z=3}$; even at $z$$\sim$4 we
reach 2 mag fainter than $M^*$.  The KDF have an area of 169
arcmin$^2$ divided into three spatially independent regions on the sky
that allow us to monitor the effects of cosmic variance.  The $U_n G
{\cal R} I$ filters and color-color selection commonality with the
work of Steidel et al.\ lets us confidently select very faint,
sub-$L^*_{z=3}$ galaxies at $z$$\sim$2, 3, and 4 by relying on their
extensive spectroscopic characterization of sample selection.  It also
allows us to confidently combine our data with theirs, thereby for the
first time consistently spanning such a large range in galaxy
luminosity at high redshift.

Our $z$$\sim$5 work is based on very deep $VI_cz'$ imaging of two
Subaru Suprime-Cam fields presented in \citet{sawicki:iwata2006,
sawicki:iwata2007}. These data are wider and deeper than earlier work
we presented in \citet{sawicki:iwata2003}: they now span a total of
1,300 arcmin$^2$ divided over two fields with $z'_{lim}(AB)$ = 26.5
and 25.5.  The large area covered by these data is essential for
determining the abundance of the rare, luminous objects, while at the
same time the depth of the data gives us the necessary ability to
probe below $L^*$ in the LF. We use $VI_cz'$ Lyman break selection to
select $z$$\sim$5 galaxies and our ongoing spectroscopic program has
already confirmed a number of $z$$\sim$5 galaxies, validating our
$z$$\sim$5 color-color selection technique
\citep{sawicki:ando2004, sawicki:iwata2007}.

\subsection{The differentially evolving luminosity function}

We find that the luminosity function of high-$z$ galaxies undergoes
evolution that is differential with luminosity. As Fig.~1(a) shows,
the bright end of the LF remains virtually unchanged, but the faint
end evolves from $z$$\sim$5 to at least $z$$\sim$3.  Our tests show
that it is highly unlikely that this observed differential evolution
is due to some systematic effect such as selection bias, our modeling
of the survey volume, etc.\
\citep[for details see][]{sawicki:kdf2}.  The effect is also 
quite significant --- the statistical probability that the
$z$$\sim$4$\rightarrow$3 evolution is {\it not} differential with
luminosity is only 1.5\%, and the result is even more robust for
$z$$\sim$5$\rightarrow$3.  We conclude that the {\it differential,
luminosity-dependent evolution} is very likely real.  If so, it must
reflect some real, intrinsic evolutionary differences between faint
and luminous LBGs.  Understanding what causes these differences will
give us insights into what drives galaxy evolution at high redshift.

\subsection{Luminosity and star formation rate densities}

The importance of the LF's faint end extends into measurements of the
cosmic luminosity density and the cosmic star formation rate that is
often derived from it.  For reasonable faint-end slopes of the LF the
bulk of the luminosity density resides in galaxies that are fainter
than $L^*$ and it is only with deep, multi-field, large-area surveys
such as ours that it is possible to reliably measure the contribution
of this dominant population.

Figure~1(b) shows the evolution of the UV luminosity density using
data from GALEX \citep{sawicki:arnouts2005, sawicki:wyder2005}, Keck
Sawicki \& Thompson (2006b), 
and Subaru \citep{sawicki:iwata2007}.
The {\it total} UV luminosity density, i.e., luminosity density due to
galaxies of {\it all} luminosities, rises from early epochs,
$z$$\geq$5, experiences either a plateau or a broad peak at
$z$$\sim$3--1, and then a decline to $z$=0.  This behavior of the
total luminosity density is dominated not by luminous galaxies, but by
sub-$L^*_{z=3}$ objects.  Indeed, it is galaxies within the rather
narrow luminosity range $L$=(0.1--1)$L^*_{z=3}$ that dominate at high
redshift, $z$$\sim$2--5, and it is the rapid evolution of this
population that drives the evolution in the {\it total} UV luminosity
density.  Thus far, this faint population has been largely neglected
in high-$z$ follow-up studies which have focused on
$L$~$^>_\sim$~$L^*_{z=3}$ LBGs.  In contrast to $z$$\geq$2, at lower
redshifts, $z$$\leq$1, the total UV luminosity density is dominated by
{\it very} faint galaxies with $L$$<$0.1$L^*_{z=3}$. The
$L$=(0.1--1)$L^*_{z=3}$ galaxies that dominated at $z$$\geq$2 are
still important but no longer dominant.  This switch in the luminosity
of the galaxies that dominate the total luminosity density is
reminiscent of galaxy downsizing, although the effect we see so far is
a downsizing in UV luminosity rather than galaxy mass.

\begin{figure}
\label{LF.fig}
\plotone{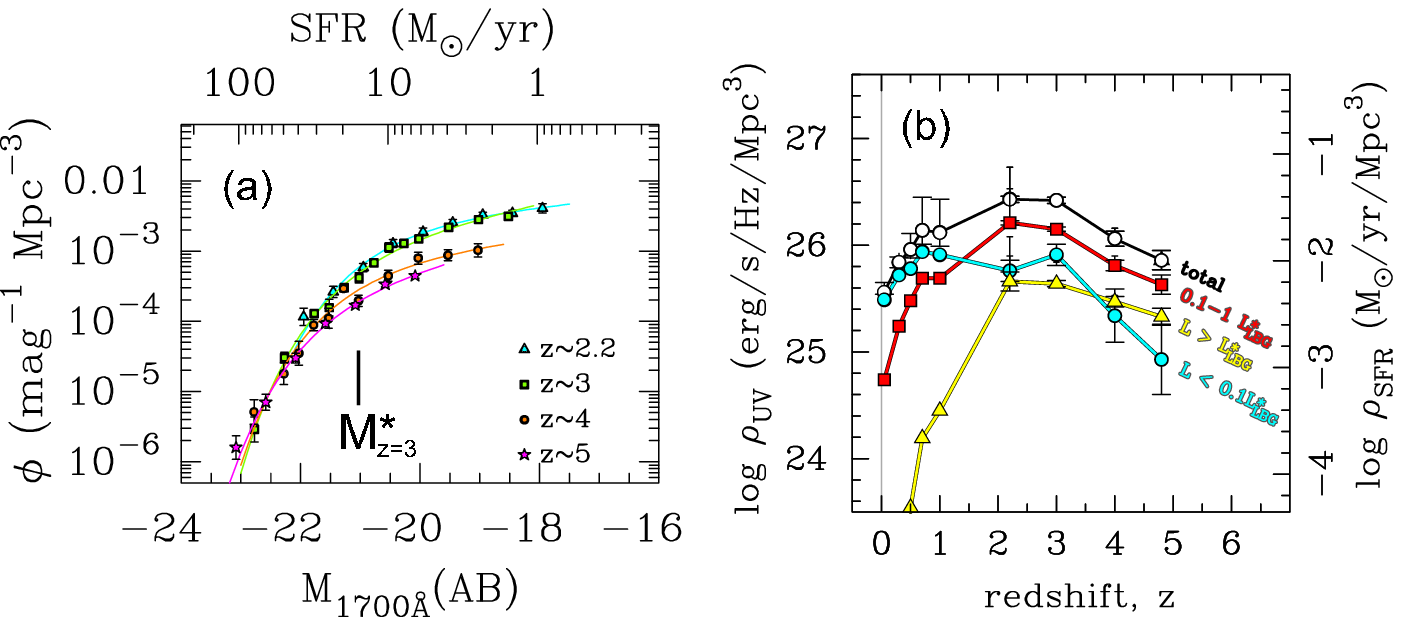}
\caption{
{\bf (a)} Luminosity-dependent evolution of the UV luminosity
function.  The $z$$\sim$5 results are from
\citet{sawicki:iwata2007} while the $z$$\sim$4, 3, and 2 points are from the
KDF work of \citet{sawicki:kdf2} and include the data of \citet{sawicki:steidel1999}
at the bright end.  The $z$$\sim$2.2 LF may be subject to a systematic
bias that {\it could} be suppressing the number density shown by up to
a factor of two; however, the $z$$\sim$3 and $z$$\sim$4 points are
{\it extremely} robust, as discussed in \citet{sawicki:kdf2}.  {\bf (b)} The
UV luminosity density of the Universe and the corresponding star
formation rate density (uncorrected for dust).  Note that the
divisions by luminosity are defined on the basis of the {\it
non}-evolving $z$$\sim$3 $L^*_{z=3}$. The $z$$\sim$5, 4, 3, and 2
points are from our work, while the lower-$z$ data are from GALEX.
The bulk of the UV luminosity density at high redshift, $z$$>$2, comes
from the faint, thus far largely neglected galaxies in the range
$L$=(0.1--1)$L^*_{z=3}$.  }
\end{figure}

\section{Physical properties from SED fitting of sub-$L^*$ galaxies at $z$$\sim$2}

\subsection{The galaxy sample and SED fitting technique}

As we have discussed above, sub-$L^*$ LBGs dominate over their more
luminous cousins, but evolve differently from them as a population.
We now wish to know what makes them {\it physically} different.  We
attempt to address this issue through broadband spectral energy
distribution (SED) studies.  Unlike in the LF work, here we do not
need large samples of objects but instead must have deep
multiwavelength photometry that spans from the UV to beyond the Balmer
break in the rest frame.  We need to go particularly deep because we
wish to study galaxies significantly fainter than $L^*$ at $z$$>$2.
No spectroscopic samples of such very faint LBGs currently exist and so we
must use photometric redshifts.

Our first foray into this area is at $z$$\sim$2 and uses the HST
WFPC2+NICMOS $U_{300}B_{450}V_{606}I_{814}J_{110}H_{160}$ data of the
northern Hubble Deep Field (HDF).  These data are ideal for us as they
are very deep, include the $U$-band imaging essential for selecting
$z$$\sim$2 LBGs, and span the age-sensitive Balmer break at the target
epoch.

After smoothing the multiband images to a common PSF and obtaining
SExtractor \citep{sawicki:bertinarnouts1996} photometry, we constructed a
$U$-band drop-out sample using the LBG color-color selection criteria
defined by 
Steidel et al,\ (1996)
for the HDF.  We then applied a
photometric redshift cut, 1.8$\leq$$z_{phot}$$\leq$2.6, where the
photometric redshifts are determined as part of the SED-fitting
procedure (see below). The lower limit of this photo-$z$ cut
eliminates potential low-$z$ interlopers and the higher limit excludes
high-$z$ LBGs for which the
$U_{300}B_{450}V_{606}I_{814}J_{110}H_{160}$ filter set does not span
the crucial, age-sensitive Balmer break.  This procedure (two-color
selection followed by a photo-$z$ cut) is very similar to the
procedure (two-color selection followed by spectroscopy) used by
\citeauthor{sawicki:steidel1996} in their work.  Our resulting sample of 
$\sim$70 objects has a mean redshift $\bar{z}_{phot}$=2.3 and in other
ways also closely mimics the BX samples of \citet{sawicki:steidel2003}
and \citet{sawicki:shapley2005}. However, it contains much fainter
galaxies, reaching down to $R(AB)$=27.\footnote{As in
\citet{sawicki:steidel1996}, the $R$ magnitude is defined via
averaging the $V_{606}$ and $I_{814}$ fluxes.}

Our SED fitting procedure follows the now well-established approach
first developed for LBGs by \citet{sawicki:sawickiyee1998} and
subsequently used by many others \citep[e.g.,][]{sawicki:papovich2001,
sawicki:shapley2001, sawicki:shapley2005, sawicki:iwata2005}.  We
compare the observed $B_{450}V_{606}I_{814}J_{110}H_{160}$\footnote
{While $U_{300}$ is essential for {\it selecting} our LBG sample, we
omit it from the SED fitting to avoid contaminating our SED analysis
with the stochastic effects of intergalactic hydrogen absorption.}
galaxy photometry with predictions of
\citet{sawicki:bruzualcharlot2003} spectral synthesis models
attenuated by \citet{sawicki:calzetti2000} dust. We fix metallicity
and star formation history and fit for burst age, star formation rate,
stellar mass, dust reddening, and redshift.  While for brighter
samples redshifts are usually constrained from spectroscopy, here we
must keep redshift as a free parameter.  However, our tests as well as
previous experience with this approach \citep{sawicki:sawicki2002,
sawicki:hall2001} show that lack of spectroscopic redshifts does not
significantly affect the results.

\begin{figure}
\label{SEDfit.fig}
\plotone{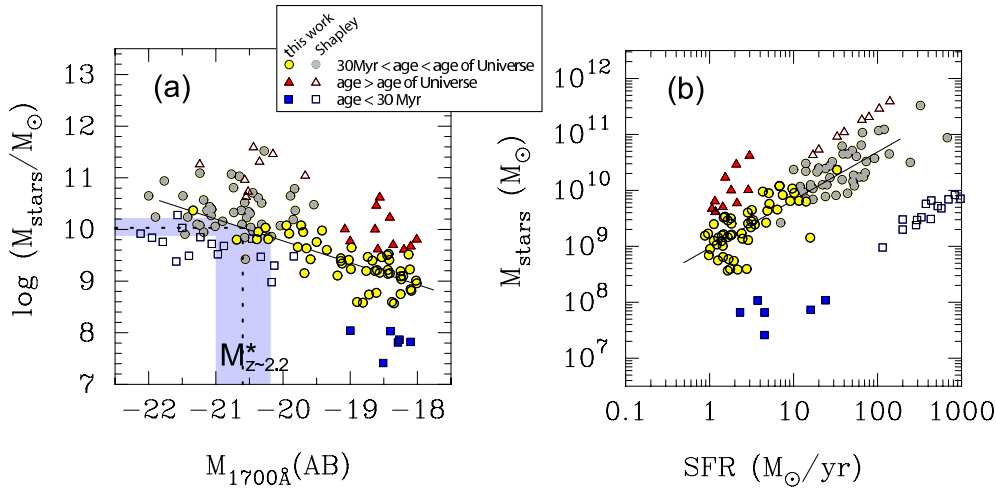}
\caption{
{\bf (a)} Stellar mass as a function of UV magnitude at
$z$$\sim$2.3.  Results for brighter galaxies fitted by 
\citet{sawicki:shapley2005} are shown along with our sample.  
Both samples were fitted with 1$Z_\odot$, constant\ SFR
\citet{sawicki:bruzualcharlot2003} models attenuated with \citet{sawicki:calzetti2000}
dust.  {\bf (b)} Stellar mass as a function of (dust-corrected) star
formation rate.  In both panels solid lines show fits to the HDF
data. }\end{figure}

\subsection{A mass-SFR relation and a characteristic mass}

Figure~2a shows that the bulk of the galaxies in our HDF sample
follows a relation between stellar mass and UV luminosity and this
relation extends into the brighter galaxies of \citet{sawicki:shapley2005}.
There appears to be good correlations between stellar mass and UV
luminosity (Fig.~2a) and stellar mass and SFR (Fig.~2b).  Outliers in
both samples turn out to be galaxies with constant-SFR best-fit ages
apparently older than the age of the Universe at $z$$=$2.3 (triangles)
and galaxies with very young ages $age_{fit}$$<$30Gyr (squares).
However, most objects show a fairly tight correlation which can be
described by the relation
$\log(M_{stars}/M_\odot)=9.0+0.86\log[SFR/(M_\odot yr^{-1})]$ (solid
line in Fig.~2b). The inclusion of the outliers in the fit does not
significantly affect this result.
 
The $L^*$ knee in the $z$$\sim$2.2 LF
($M^*_{1700A}$=$-20.6$ at $z$$\sim$2.2: Sawicki \& Thomp\-son, 2006a)
corresponds to a stellar mass
$M_{stars}(L^*)$$\sim$10$^{10}$$M_\odot$ (Fig~2a). This $z$$\sim$2
characteristic mass is remarkably similar to the
3$\times$10$^{10}$$M_\odot$ characteristic mass that at low redshift
marks the transition between the two distinct galaxy ``bimodality''
families \citep{sawicki:kauffmann2003}.  This similarity of
characteristic masses suggests that the processes \citep[see,
e.g.,][]{sawicki:dekel2006} responsible for the observed galaxy
bimodality at low $z$ may have also been operating at $z$$\sim$2.
Moreover, the near-unity slope of the logarithmic relation between
stellar mass and UV luminosity (Fig.~2a) suggests that the bulk of the
detectable stellar mass in $z$$\sim$2.2 LBGs resides in the thus-far
poorly studied galaxies with $L^*$$>$$L$$>$0.1$L^*$, just as does the
bulk of the total UV luminosity density.

The outliers with extremely young fit ages (square symbols) are
apparently either undermassive for their SFRs or overluminous for
their masses.  In the former case they may move up onto the mass-SFR
relation as they age; in the latter they could be experiencing strong
star-forming bursts that dramatically elevate their normally low
luminosities.  Outliers with very old best-fit ages (triangles), are
also consistent with variable star formation histories: constant star
formation histories tend to yield the largest ages
\citep[e.g.,][]{sawicki:sawickiyee1998,sawicki:papovich2001} and thus the old ages
here may simply be telling us that star formation in these galaxies
was more intense in the past.  All these scenarios advocate that at
least some $z$$\sim$2 LBGs, both faint and luminous, have variable
star formation histories --- consistent with one of the scenarios
proposed by \citet{sawicki:kdf2} to explain the evolving LBG luminosity
function.

\section{Concluding remarks}

The thus-far largely unexplored sub-$L^*$ LBGs dominate the UV
luminosity density of the universe at $z$$>$2.  Moreover, their
stellar masses appear to correlate with UV luminosities down to very
faint magnitudes at least at $z$$\sim$2 and so, just as they dominate
the UV luminosity density, these objects may also contain a
substantial fraction of the formed stellar mass by that redshift.  At
the same time, however, the differentially-evolving luminosity
function suggests that these sub-$L^*$ LBGs differ substantially from
their more luminous cousins.  It is not yet clear what mechanism is
behind this differential evolution, but the bursty nature of at least
some of the $z$$\sim$2 galaxies means that changes in the timescales
of star-burst episodes may play a role. We will address this and other
issues as we extend our SED-fitting and other differential studies to
higher redshifts.



\end{document}